\documentclass[aip,pof,reprint]{revtex4-1}

\usepackage{graphicx}
\usepackage{amsmath}
\usepackage{amssymb}
\usepackage{bm}
\usepackage{dcolumn}
\usepackage{comment}

\usepackage[utf8]{inputenc}
\usepackage[T1]{fontenc}
\usepackage{etoolbox}
\usepackage{float}
\usepackage{multirow}

\usepackage{ulem}
\usepackage{xcolor}

\makeatletter
\def\@email#1#2{%
 \endgroup
 \patchcmd{\titleblock@produce}
  {\frontmatter@RRAPformat}
  {\frontmatter@RRAPformat{\produce@RRAP{*#1\href{mailto:#2}{#2}}}\frontmatter@RRAPformat}
  {}{}
}%
\makeatother

\usepackage[ruled,linesnumbered]{algorithm2e}
\usepackage{hyperref}

\begin{document}

\title{Solving Euler equations with Multiple Discontinuities via Separation-Transfer Physics-Informed Neural Networks}

\author{Chuanxing Wang}
    \affiliation{Graduate School, China Academy of Engineering Physics, Beijing 100088, China}
    
\author{Hui Luo}
 \email{hluo@gscaep.ac.cn}
 \affiliation{Graduate School, China Academy of Engineering Physics, Beijing 100088, China}

\author{Kai Wang}

\affiliation{Zhejiang Institute of Modern Physics, School of Physics, Zhejiang University, Hangzhou Zhejiang 310058, China}
 
\author{Guohuai Zhu}
    \affiliation{Beijing Computational Science Research Center, Beijing 100193, China}

\author{Mingxing Luo}
    \affiliation{Beijing Computational Science Research Center, Beijing 100193, China}

\date{\today}

\begin{abstract}
Despite the remarkable progress of physics-informed neural networks (PINNs) in scientific computing, they continue to face challenges when solving hydrodynamic problems with multiple discontinuities. In this work, we propose Separation-Transfer Physics-Informed Neural Networks (ST-PINNs) to address such problems. By sequentially resolving discontinuities from strong to weak and leveraging transfer learning during training, ST-PINNs significantly reduce the problem complexity and enhance solution accuracy. To the best of our knowledge, this is the first study to apply a PINNs-based approach to the two-dimensional unsteady planar shock refraction problem, offering new insights into the application of PINNs to complex shock–interface interactions. Numerical experiments demonstrate that ST-PINNs more accurately capture sharp discontinuities and substantially reduce solution errors in hydrodynamic problems involving multiple discontinuities.
\end{abstract}

\pacs{}

\maketitle 

\section{INTRODUCTION}
PINNs have been widely applied to solve diverse problems in fluid mechanics \cite{Raissi, Rsissi2, qiu1, zhao1}, solid mechanics \cite{HAGHIGHAT, HAGHIGHAT2}, and plasma physics \cite{wang1, Oh}.
By embedding partial differential equations (PDEs) along with their initial and boundary conditions into the loss function, PINNs convert the task of solving PDEs into an optimization problem.
PINNs is a machine learning method that seamlessly combines data and physical knowledge. It has notable advantages in addressing problems with sparse data, such as flow field reconstruction \cite{Rsissi2, Xu1, Peng1}, and overcomes the limitations of purely data-driven methods, which typically require large datasets.

Hydrodynamic problems involving discontinuities, such as shock propagation and shock-interface interactions, are fundamentally significant in fields such as astrophysics \cite{muller,kifonidis}, inertial confinement fusion \cite{atzeni,betti}, and aerospace engineering \cite{yangA,marble}. The presence of shock profoundly affects the entire flow field, highlighting the necessity of resolving discontinuous hydrodynamic problems. Unfortunately, PINNs face significant challenges in solving problems involving discontinuities, primarily due to overly large loss function values around discontinuities, which adversely impact the optimization process. This phenomenon, known as gradient pathology \cite{Sifan}, typically causes PINNs to enter failure modes.

To enhance its accuracy and robustness in addressing discontinuous problems, various variants of PINNs have been proposed.
The first type of approach is increasing the sampling density near discontinuities to enhance their influence during the training process, enabling PINNs to more accurately capture discontinuities. The feasibility of this approach has been demonstrated by solving a one-dimensional problem with a moving contact discontinuity and a two-dimensional steady problem with an oblique shock \cite{MAO1}.
DeepXDE \cite{Lu1} uses the absolute value of the loss function to approximate the location of discontinuities, enabling adaptive resampling near discontinuities during training. However, this approach does not universally mitigate gradient pathologies and performs well only for relatively simple problems.
The second, more effective approach is domain decomposition, which mitigates gradient pathologies during training by reducing the number of discontinuities in each solution subdomain. CPINNs \cite{JAGTAP1} divides the solution domain into subdomains and introduces flux conservation constraints into the loss functions at their interfaces. Additionally, physical quantities at interfaces are averaged. However, domain decomposition can still result in significant errors near discontinuities. Derived from CPINNs, XPINNs \cite{JAGTAP2} extends domain decomposition to both spatial and temporal dimensions. The domain decomposition approach has also been applied in PINNs to solve diffusion problems \cite{YAO}.
The third effective approach is to consider weights in the PDEs to reduce the adverse effects of discontinuities during training. 
Specifically, the PINNs-WE \cite{LiLiu} method uses a velocity gradient-based weighting scheme to suppress overly large loss function values in sampling regions near discontinuities. 
In addition, it also employs Rankine–Hugoniot (RH) relation and a global physical conservation constraint to improve the shock-capturing performance.
Those adjustments improve training performance in smooth regions while also ensuring effective training in discontinuous regions. 
GA-PINNs extends the weighting scheme of PINNs-WE to include a combination of weights for density gradients, velocity gradients, and pressure gradients\cite{Antonio}.
This approach effectively mitigates the gradient pathology and performs well in problems with a single strong discontinuity, such as the one-dimensional Sod problem. 
These three aforementioned methods collectively improve the performance of PINNs in solving problems with discontinuities.

In practical science and engineering applications, problems involving multiple discontinuities are frequently encountered. 
The gradient imbalance across different discontinuous regions makes these problems highly challenging, often resulting in that some discontinuities are well trained, while others are not.
The three methods mentioned above are difficult to achieve ideal results when solving this kind of problem, especially those with severe gradient imbalance.
Thus, effective improvements are required.

This paper proposes the Separate-Transfer Physics-Informed Neural Networks (ST-PINNs) approach to address such challenges more effectively. 
This approach first uses a variant method of PINNs to solve problems with multiple discontinuities, of which the strongest discontinuity can be well captured, and then a 
first-round trained model is generated.  
Then, the well-captured discontinuity divides the solution domain into two subdomains.
In each subdomain, the physics values predicted by the first-round trained model near the discontinuity are used as boundary conditions. 
Transfer learning is then applied using the first-round trained model to improve the prediction performance of poorly trained discontinuity regions, effectively mitigating the gradient pathology encountered during the training process. 
Through iterative region separation and transfer learning, regions containing all discontinuities 
in principle can be well captured in sequence. 
Finally, by combining 
{\color{black} results} of different subdomains, the predictions for the entire solution domain are obtained. 
This approach has been validated in one-dimensional shock-interface interaction problem, quasi-one-dimensional planar shock-interface interaction problem, and two-dimensional unsteady  planar shock refraction problem. 
Notably, to demonstrate the effectiveness of the idea of the domain separation and transfer learning, the numerical experiments in this study rely exclusively on PDEs and their initial conditions, without incorporating data from traditional numerical simulations or experiments. 

The organization of this paper is as follows:  
Section \ref{section2} introduces the Euler equations for multiple discontinuities problems and the fundamental framework of PINNs, discusses its failure modes of the problems with multiple discontinuities, and analyzes gradient pathology in such cases.  Section \ref{section3} provides a detailed description of the specific steps of ST-PINNs. 
Section \ref{section4} demonstrates the effectiveness of ST-PINNs by solving typical multiple discontinuities problems. 
Finally, Section \ref{section5} discusses and concludes this paper.

\section{BACKGROUND}\label{section2}

\subsection{Euler Equations}
Shock-related  problems represent a typical class of multi-discontinuity problems, which are prevalent in fields such as hydrodynamics, astrophysics, and high-energy-density physics. Ideal hydrodynamics problems involving shocks are typically described by the Euler equations, which neglect viscosity. To validate the capability of ST-PINNs in addressing such problems, we applied it to one-dimensional shock-interface interaction problem, quasi-one-dimensional planar shock-interface interaction problem, and two-dimensional planar shock refraction problem by solving the conservative form of one-dimensional and two-dimensional Euler equations. The Euler equations consist of the density equation, momentum equation, and energy equation, and are closed by the equation of state (EOS), which links the fluid pressure to the density and energy.

For a one-dimensional problem, the governing equations are shown as Eq. (\ref{eq1})
\begin{equation}
    \frac{\partial U}{\partial t}+\frac{\partial F}{\partial x}=0,
    \label{eq1}
\end{equation}
where,
\begin{equation}
    U=\begin{pmatrix}
 \rho\\
 \rho u\\
E \
\end{pmatrix},\quad
F=\begin{pmatrix}
 \rho u\\
 \rho u^2+p\\
u(E+p)
\end{pmatrix}.
\label{eq2}
\end{equation}
Here, $U$ is the vector of conserved variables, which includes the mass density $\rho$, momentum density $\rho u$, and total energy density $E$. 
$F$ represents the flux vector in the $x$ direction. 
For a polytropic gas, EOS is $ E=\frac{1}{2}\rho u^2+\frac{p}{\gamma -1}$, where $u$ is the velocity, $p$ is the pressure, and $\gamma$ is an adiabatic exponent. 

For a two-dimensional problem, Euler equations are expressed in the form of Eq. (\ref{eq3}).

\begin{equation}
    \frac{\partial U}{\partial t}+\frac{\partial F}{\partial x} + \frac{\partial G}{\partial y}=0,
    \label{eq3}
\end{equation}
where
\begin{equation}
    U=\mathbf{\begin{pmatrix}
 \rho \\
\rho u \\
 \rho v\\
E
\end{pmatrix}} ,\quad
    F=\begin{pmatrix}
 \rho u\\
 \rho u^2+p\\
\rho u v\\
u(E+p)
\end{pmatrix},\quad
G=\begin{pmatrix}
 \rho v\\
\rho u v\\
 \rho v^2+p\\
v(E+p)
\end{pmatrix}. 
\label{eq4}
\end{equation}
Here, $F$ and $G$ are the flux vectors in the $x$ and $y$ directions, respectively.  
$u$ and $v$ are the velocity components in the $x$ and $y$ directions. 
For a polytropic gas, EOS is $E=\frac{1}{2}\rho (u^2+v^2)+\frac{p}{\gamma -1}$.

The adiabatic exponent is set to 1.4 for all the cases in this paper.

\subsection{BASIC FRAMEWORK OF PINNs}
{\color{black} Consider} a PDE with initial and boundary conditions as shown in Eq. (\ref{eq:PINN}).
\begin{equation}
    \begin{cases}
  {\mathcal V}_{t}+\mathcal{O}[{\mathcal V}]=0, & (\bm{x},t) \in \bm{\Omega} \times[0,\rm{T}];  \\
  {\mathcal V}(\bm{x} ,0)={\mathcal H}(\bm{x}), & \bm{x} \in \bm{\Omega}; \\
  \mathcal{B}[{\mathcal V}(\bm{x} ,t)]=0, & (\bm{x},t) \in \partial \bm{\Omega} \times[0,\rm{T}].  
\end{cases}
    \label{eq:PINN}
\end{equation}
Here, ${\mathcal V}$ is the quantity to be solved, and $\bm{x}$ and $t$ denote spatial and temporal coordinates, respectively. 
${\mathcal V}_t$ represents the partial derivative of ${\mathcal V}$ with respect to time and $\mathcal{O}$ is a nonlinear operator that represents spatial differentiation. 
The spatial domain is $\bm{\Omega} \subset \mathbb{R}^d$, with its boundary denoted by $\partial \bm{\Omega}$. The initial condition is $\mathcal{H}$, and $\mathcal{B}$ is the boundary operator, which specifies boundary conditions such as Dirichlet, Neumann, and periodic conditions.

PINNs solve PDEs by converting them into optimization problems, where the solution is obtained by minimizing the loss function. 
{\color{black} In this work, a fully connected neural network with $\rm{L}$ hidden layers is used as shown in Eq. (\ref{eq:hidden_layer})}

\begin{equation}
     \mathcal{N}^{l} =\sigma (W^l \mathcal{N}^{l-1} +b^l) \quad 1\le l\le \rm{L}+1,
     \label{eq:hidden_layer}
\end{equation}
where $\mathcal{N}^{0}$ is input layer, $\mathcal{N}^{\rm{L}+1}$ is output layer, $\sigma$ is activation function, {\color{black} $W$ and $b$ are weights and biases respectively}. 
The relationship between the output layer and input layer can be expressed as Eq. (\ref{eq:output_input}).

\begin{equation}
    \hat{\mathcal{V}}(\bm{x}, t)=\mathcal{N}(\bm{x},t;\theta )= \mathcal{N}^{\rm{L}+1} \circ  \mathcal{N}^{\rm{L}} \circ \cdots \circ  \mathcal{N}^{0}(\bm{x},t) 
    \label{eq:output_input}
\end{equation}
with $\theta$ collectively refers to weights and biases.

The total loss function $L_{\rm{sum}}$ of PINNs consists of three components: the loss function of the initial condition $L_{\rm{IC}}$, the loss function of the boundary condition $L_{\rm{BC}}$, and the residual loss function $L_{\rm{F}}$, along with their corresponding weights, as shown in Eq. (\ref{eq:LOSSFUNCTION}). 
\begin{equation}
    \begin{aligned}
    &L_{\rm{sum}}=\omega_{\rm{IC}}L_{\rm{IC}}+\omega_{\rm{BC}}L_{\rm{BC}}+\omega_{\rm{F}}L_{\rm{F}},  \\
    &L_{\rm{F}}=\frac{1}{\rm{N}_{\rm{F}}}\sum_{i=1}^{\rm{N}_{\rm{F}}} \left|\frac{\partial {\mathcal V}_{\rm{NN}}\left(\bm{x} _{i}^{\rm{F}},t_i^{\rm{F}}\right)}{\partial t}+\mathcal{O}\left[{\mathcal V}_{\rm{NN}}\left(\bm{x} _{i}^{\rm{F}},t_i^{\rm{F}}\right)\right]\right|^2,   \\
    &L_{\rm{IC}}=\frac{1}{\rm{N}_{\rm{IC}}}\sum_{j=1}^{\rm{N}_{\rm{IC}}} \left|{\mathcal V}_{\rm{NN}}\left(\bm{x} _{j}^{\rm{IC}},0\right)-{\mathcal H}\left(\bm{x}_{j}^{\rm{IC}},0 \right)\right|^2,\\
    &L_{\rm{BC}}=\frac{1}{\rm{N}_{\rm{BC}}}\sum_{k=1}^{\rm{N}_{\rm{BC}}} \left|\mathcal{B}\left[{\mathcal V}_{\rm{NN}}\left(\bm{x} _{k}^{\rm{BC}},t_k^{\rm{BC}}\right)\right]\right|.
    \end{aligned}
    \label{eq:LOSSFUNCTION}
\end{equation}
Here,
$\omega_{\rm{IC}}$, $\omega_{\rm{BC}}$ and $\omega_{\rm{F}}$ are weights of the initial condition, the boundary condition, and the residual loss function, respectively. 
The term ${\mathcal V}_{\rm{NN}}(\bm{x}, t)$ denotes predictions of the neural network at spatial coordinate $\bm{x} \in \bm{\Omega}$ and temporal coordinate $t \in [0, T]$.
$\mathrm{N}_\mathrm{F}$ sampling points $(\bm{x}_{i}^{\mathrm{F}}, t_{i}^{\mathrm{F}})$ are selected in the solution domain $\bm{\Omega} \times [0, T]$. 
$\mathrm{N}_{\mathrm{IC}}$ points $(\bm{x}_{j}^{{\rm{IC}}}, 0)$ are selected at the initial time $t = 0$ within $\bm{\Omega}$, and $\rm{N}_{\rm{BC}}$ points $(\bm{x}_{k}^{\rm{BC}}, t_{k}^{\rm{BC}})$ are selected on the boundary $\partial \bm{\Omega} \times [0, T]$.
By minimizing the loss function, PINNs obtain suitable neural network parameters $\theta$ to predict the solution of the PDEs.

\subsection{Failure Mode of PINNs}

The effectiveness of PINNs is based on minimizing the loss function.
However, for problems with multiple discontinuities, two types of gradient imbalances can occur: imbalance among the gradients of $L_\mathrm{F}$, $L_{\mathrm{IC}}$ and $L_{\mathrm{BC}}$ with respect to $\theta$, and imbalance among the gradients of the loss function in different regions.

For the first type of imbalance, if the weights of the loss functions are set equal, i.e., $\omega_{\mathrm{IC}}=\omega_{\mathrm{BC}}=\omega_{\mathrm{F}}=1$, 
the training process is affected when one loss function term is excessively large compared to the others. 
The largest loss term dominates the training process, directly affecting the optimization of other terms. 
To mitigate this kind of gradient pathology, an adaptive weighting strategy \cite{Sifan} can be applied. By adaptively adjusting the weights $\omega_{\rm{IC}}$ and $\omega_{\rm{BC}}$ 
at different stages of training, the gradients of $L_{\rm{F}}$, $L_{\rm{IC}}$, and $L_{\rm{BC}}$ can be balanced, ensuring that all loss terms are effectively optimized.
This strategy calculates the transition weights $\hat{\omega_i}$ after a specified number of steps using Eq. (\ref{eq:hat_omega}) as

\begin{equation}
        \hat{\omega _i}=\frac{{\rm{max}} \left\{|\nabla_{\theta}L_{\rm{F}}(\theta_n)|\right\} }{\overline{|\nabla_{\theta}\omega_i L_i(\theta_n)|} },\quad i=\rm{IC}\,\, \rm{or}\,\, \rm{BC},\quad \\
        \label{eq:hat_omega}
\end{equation}
where $\theta_n$ is the neural network parameters at the $n$-th iteration, and $\nabla_{\theta}$ is the gradient with respect to $\theta$, specifically referring to the element-wise partial derivative of each weight $W$, resulting in a vector. 
$|\cdot|$ represents the element-wise absolute value, and $\overline{|\cdot|}$ indicates the mean of the element-wise absolute values.
Subsequently, the weight of the loss term with the smaller absolute value from the previous step increases, while that of the larger term decreases. 
The adaptive weight $\omega_i$ is then updated according to Eq. (\ref{eq:omega}) as
\begin{equation}
    \omega_i=(1-\alpha)\omega_i+\alpha \hat{\omega_i}
    \label{eq:omega}
\end{equation}
with $\alpha$ a tunable hyper-parameter, thereby achieving a balance among different loss functions.

Another type of gradient imbalance occurs when the loss functions of sampling points in different regions vary significantly. 
Specifically, the loss function values of sampling points near discontinuities are notably larger than those in smooth regions, causing the former to dominate the training process.
To deal with this problem, PINNs-WE \cite{LiLiu} identifies the location of the discontinuity by a velocity-dependent weight $\lambda$, 
as
\begin{equation}
    \begin{aligned}
    &\lambda =\frac{1}{k(|\nabla \cdot \bm{u}| - \nabla \cdot \bm{u})+1  },
    \end{aligned}
    \label{eq:lambda}
\end{equation}
and its residue loss function is then
\begin{equation}
 L_{\rm{F}}=\frac{1}{\rm{N}_{\rm{F}}}\sum_{i=1}^{\rm{N}_{\rm{F}}} \lambda |\frac{\partial {\mathcal V}_{\rm{NN}}(\bm{x} _{i}^{\rm{F}},t_i^{\rm{F}})}{\partial t}+\mathcal{O}[{\mathcal V}_{\rm{NN}}(\bm{x} _{i}^{\rm{F}},t_i^{\rm{F}})]|^2,
 \label{eq:loss_F_WE}
\end{equation}
where $k$ in Eq. (\ref{eq:lambda}) is an adjustable hyper-parameter.
In addition, the Rankine–Hugoniot relation is introduced, as well as a global physical conservation constraint, to improve the shock-capturing performance. 
The similar method, GA-PINNs\cite{Antonio}, extends the local weighting scheme to include a combination of density gradients, velocity gradients,
and pressure gradients. 
In particular, the local weight in one-dimensional problem is 
\begin{equation}
    \lambda = \frac{1}{1+\sum_{i=1} \alpha_i |\partial_x {\mathcal V}_{{\rm NN},i}|^{\beta_i}},
\end{equation}
where $i$ runs over all predictive variables, $\alpha_i$ and $\beta_i$ are adjustable hyper-parameters.

These methods show a significant improvement in some discontinuity computing examples compared to traditional PINNs.
However, for problems with multiple discontinuities, particularly those that exhibit significant differences in discontinuity intensity, 
these methods capture stronger discontinuities better than weaker discontinuities, thus they are not sufficient to accurately capture all discontinuities and achieve perfect results. 
Advanced strategies are required to mitigate the gradient pathology among different discontinuities.

\section{METHODOLOGY}\label{section3}

From Euler equations, it is clear that the spatial derivatives of physical quantities theoretically approach infinity at discontinuity, making the problem difficult to solve. 
We propose {\color{black} the separation-transfer physics-informed neural networks (ST-PINNs)} method to address the gradient imbalance problem across different regions of the entire solution domain for hydrodynamic problems involving multiple discontinuities.

Discontinuity computing is challenging for both traditional numerical methods and PINNs, often leading to failure modes.
To distinguish between different discontinuities within the solution domain, $I\equiv \frac{|{\mathcal V}_{+}-{\mathcal V}_{-}|}{\Delta \bm{x}}$ is defined to quantify the discontinuity intensity.  
Here, as mentioned in last section, $``\mathcal V"$ represents different physical quantities (e.g., density, velocity, pressure), and $\Delta \bm{x}$ is the grid width at the discontinuity.
The subscripts {\color{black} $``+"$ and $``-"$ denote the regions at different sides of the discontinuity.} 
A larger $I$ signifies a stronger discontinuity, with the strongest one in the solution domain being referred to as the primary discontinuity. 
The presence of multiple discontinuities can cause the training process to enter a failure mode, in practice, the primary discontinuity dominates the training process, resulting in suboptimal training for the other discontinuities. 
The main steps of ST-PINNs are outlined in Algorithm. \ref{algorithm:ST-PINNs}. 

\begin{figure}
\centering
\includegraphics[scale=1.0]{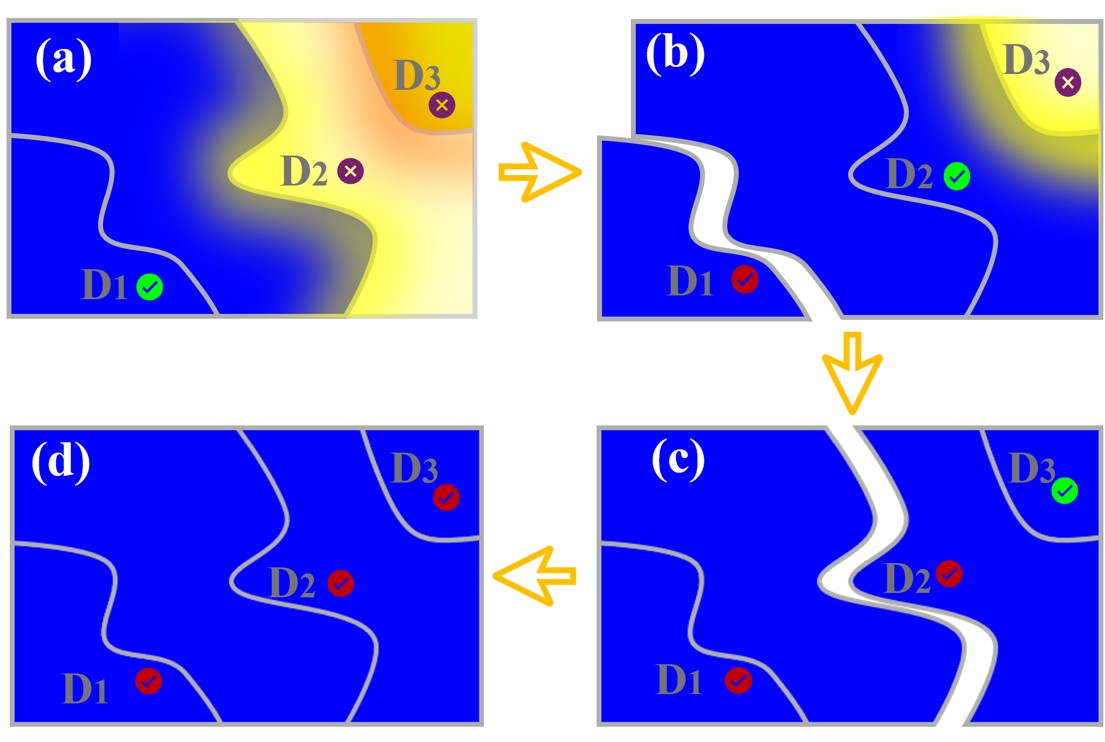}
\caption{Schematic of ST-PINNs. (a) Solve the entire solution domain by variants of PINNs, generating Model 1, with only the strongest discontinuity D1 well-captured. (b) Identify the location of discontinuity D1. The solution domain is divided into subdomains S1 and S2 along D1, with D2 becoming the primary discontinuity in S2. Transfer learning is performed based on Model 1 to generate Model 2, with which D2 is well-captured. 
(c) Identify the location of discontinuity D2. The subdomain S2 is further divided into S3 and S4 at D2, with D3 becoming the primary discontinuity in S4. Transfer learning is performed based on Model 2 to generate Model 3, with which D3 is well-captured.
(d) Models 1, 2, and 3, which predicts different subdomains well, are combined to obtain the predictions for the entire solution domain.}
\label{fig:ST-PINN}
\end{figure}

The algorithm is schematically illustrated by FIG. \ref{fig:ST-PINN}. 
As shown in FIG. \ref{fig:ST-PINN}(a), suppose there are three discontinuities in the entire solution domain, D1, D2, and D3. If the discontinuity D1 is stronger than D2 and D3, D1 is considered the primary discontinuity. During the training process, the loss function at sampling points near D1 dominates, causing the model to be well trained only for the primary discontinuity D1, while the other discontinuities remain poorly trained. 
This gradient imbalance among multiple discontinuities results in failure modes.

\begin{widetext}
\begin{algorithm*}[H]
\SetAlgoLined 

\caption{\textbf{\textrm{Separation-Transfer Physics-Informed Neural Networks}}} 
\label{algorithm:ST-PINNs}

Train via variants of PINNs until the strongest discontinuity can be well captured\;    

Output the model parameters $\theta_i$ and predictions ${\mathcal V}_{\rm NN}$(e.g. $\rho_{\rm{NN}}$, $u_{\rm{NN}}$, $p_{\rm{NN}}$)\;

\While{\textrm{existing poorly predicted discontinuity regions 
}}
{Compute 
$I$ to capture the primary discontinuity position $\bm{r}_{\rm{disc}}$\;

Fit discontinuity location $\bm{r}_{\rm{disc}}$ ($t$)\;

Obtain ${\mathcal V}_{\rm NN,\pm}$ on both sides of $\bm{r}_{\rm{disc}}$\;

Divide the solution domain into two subdomains from $\bm{r}_{\rm{disc}}$\;


Apply transfer learning based on the previous model to evaluate each subdomain with poorly predicted discontinuities by using ${\mathcal V}_{\rm NN,\pm}$ as the boundary conditions of the subdomain\;

Output new model parameters $\theta_i'$ and predictions ${\mathcal V}_{\rm NN}'$\;

Evaluate whether the trained subdomains still contain poorly predicted discontinuity regions\;
}

Combine all predictions corresponding to different subdomains to obtain results for the entire solution domain.
 
\end{algorithm*}
\end{widetext}

In Step 1, ST-PINNs uses variants of PINNs to solve the problem across the entire solution domain, generating Model 1. 
However, due to the gradient imbalance, only the primary discontinuity D1 is well-captured.
In Step 2, the discontinuity strength $I$ is calculated for a chosen physical quantity (e.g., density, velocity, pressure) to identify the location of D1, which corresponds to the position with maximum $I$. 
For example, to identify the location of discontinuities using the pressure difference, the solution domain is first discretized: time is divided into N intervals 
between $\rm{t}_0$ and $\rm{T}$. At each time step, the space is also discretized, and $I$ is calculated at every grid point. 
This process determines the location of primary discontinuity and predictions on both sides of it can be obtained. 
In Step 3, the entire solution domain is divided into two subdomains S1 and S2 according to the primary discontinuity D1. As illustrated in FIG. \ref{fig:ST-PINN}(b), S2 contains poorly trained discontinuities D2 and D3. 
The predictions on the S2-adjacent side of D1 are used as boundary conditions. Transfer learning is then performed based on Model 1 to generate Model 2, ensuring that D2 is well-captured.
In Step 4, as illustrated in FIG. \ref{fig:ST-PINN}(c), if discontinuity D3 is still poorly-captured, subdomain S2 is further divided into subdomains S3 and S4 from the discontinuity D2. The predictions on the S4-adjacent side of D2 are used as boundary conditions for subdomain S4 training. 
Transfer learning is then applied in S4 based on Model 2 to generate Model 3, ensuring that D3 is well-captured.
In Step 5, if any discontinuities remain poorly trained, additional domain separation and transfer learning are conducted based on Step 4. Once all discontinuities are well-captured, the predictions from different subdomains are seamlessly combined 
to reconstruct the predictions of the entire solution domain, as illustrated in FIG. \ref{fig:ST-PINN}(d).

In the following examples, we demonstrate that regardless of the specific PINNs variant used in the initial training step, once the primary discontinuity is captured with sufficient accuracy, the ST-PINNs framework can consistently enhance the final predictions. In our implementation, the same network architecture is employed across all stages, with only the loss function weights adjusted accordingly.

\section{NUMERICAL RESULTS}\label{section4}

In this section, we apply ST-PINNs to solve shock-interface interaction problem and shock refraction problem. Since these cases simulate an infinitely large interface, the boundary conditions are not applied.

Before discussing each example in detail, we first present the same setup among them.
In these cases, the implementation of the code is based on PyTorch \cite{pytorch}.
During the training process, the optimizer is Adam and the \textrm{\texttt{tanh}} function is used as the activation function.  
The performance of neural network prediction is evaluated using the $L^2$ relative error, as defined in Eq. (\ref{eq:L2error}),
\begin{equation}
    L^2_{\rm error}=\sqrt{\frac{(f_{\rm pred}-f_{\rm true})^2}{f_{\rm true}^2}},
    \label{eq:L2error}
\end{equation}
where $f_{\rm pred}$ is the prediction and $f_{\rm true}$ is the true value.
{\color{black}In the cases under consideration, $f_{\rm true}$ is the result of traditional numerical simulations done by JAX-FLUIDS\cite{JAX-Fluids}. }

\subsection{One-dimensional shock-interface interaction}\label{section4_1}

We apply ST-PINNs to solve a one-dimensional shock-interface interaction problem. 
When an incident shock propagates from a light medium to a heavy medium, transmitted and reflected shocks are generated. 
By setting the time at which the incident shock reaches the interface as $t=0$, this problem can be equivalently considered as a one-dimensional Riemann problem, which is commonly used as a test case to validate the accuracy of algorithms and code implementations. 

The solution domain is set to $x \times t = [0.0, 1.0] \times [0.0, 0.1]$, with initial conditions specified in Eq. (\ref{eq:ShockInterface1D1}).
\begin{equation}
    \rho,u,p=\begin{cases}
    (0.919,\,0.694,\,1.833),  & \rm{if~} 0.0\le x< 0.5;
    \\
    (1.0,\,\,0.0,\,\,1.0),        &
\rm{if~}0.5\le x\le 1.0.
    \end{cases}
    \label{eq:ShockInterface1D1}
\end{equation}

\begin{table}
\centering
\caption{\label{table:Parameter1D} Hyper-parameters of ST-PINNs based on PINNs-WE method. Stage 1 refers to the epochs from 1 to 30000, Stage 2 refers to the epochs from 30001 to 50000, Stage 3 refers to the epochs from 50001 to 60000.}
\begin{tabular}
{p{4.0cm}p{2.5cm}p{1.5cm}p{0.8cm}}
    \toprule  
     Parameters & Model 1.1 & Model 1.2 \\
    \hline
     Number of layers    & 7 & 7 \\
     Neurons & 70 & 70     \\
     
     Residual Points & 10000 & 10000  \\
     Initial points & 3000 & 1000\\
     Boundary Points &NAN & 1000 \\
     Epoch & 60000 & 60000 \\
     Learning rate(Stage 1) & 1e-3 & 1e-5 \\
     Learning rate(Stage 2) & 1e-5 & 1e-6 \\
     Learning rate(Stage 3) & 1e-6 & 1e-7 \\
     $\omega_{\rm{F}}$ & 1 & 1 \\
     $\omega_{\rm{IC}}$(Stage 1-2) & 10 & 10\\
     $\omega_{\rm{IC}}$(Stage 3) & 5 & 10\\
     $\omega_{\rm{BC}}$& NAN & 10 \\
     $\omega_{\rm{RH}}$& 10 & NAN \\
     $\omega_{\rm{CON}}$& 10 & NAN \\
     
    \hline
\end{tabular}
\end{table}

In this case, PINNs-WE\cite{LiLiu} is first chosen as the training model in ST-PINNs, then, as a complementary test, GA-PINNs is applied. 

In PINNs-WE, a local weight parameter $\lambda$ as shown in Eq. (\ref{eq:lambda}) is introduced to reduce the impact of the loss function near discontinuity. 
The loss function is consist of five parts
\begin{equation}
L_{\rm{WE}}=\omega_{\rm{F}}L_{\rm{F}}+\omega_{\rm{IC}}L_{\rm{IC}}+\omega_{\rm{BC}}L_{\rm{BC}}+\omega_{\rm{RH}}L_{\rm{RH}}+\omega_{\rm{CON}}L_{\rm{CON}},
\label{eq:loss_WE}
\end{equation}
where $L_{\rm F}$ is the residual loss function as shown in Eq. (\ref{eq:loss_F_WE}), $L_{\rm{RH}}$ and $L_{\rm{CON}}$ are the introduced loss functions regarding the RH relation and global physical conservation constraint, with $\omega_{\rm{RH}}$ and $\omega_{\rm{CON}}$ corresponding global weights.
The discontinuity locations can be approximated using ${{\partial u}}/{{\partial x}}$, and the hyper-parameter $k$ adjusting local weight $\lambda$ as shown in Eq. (\ref{eq:lambda}) is set to 0.6.

\begin{figure*}[htbp]
\centering
\includegraphics[scale=0.95]{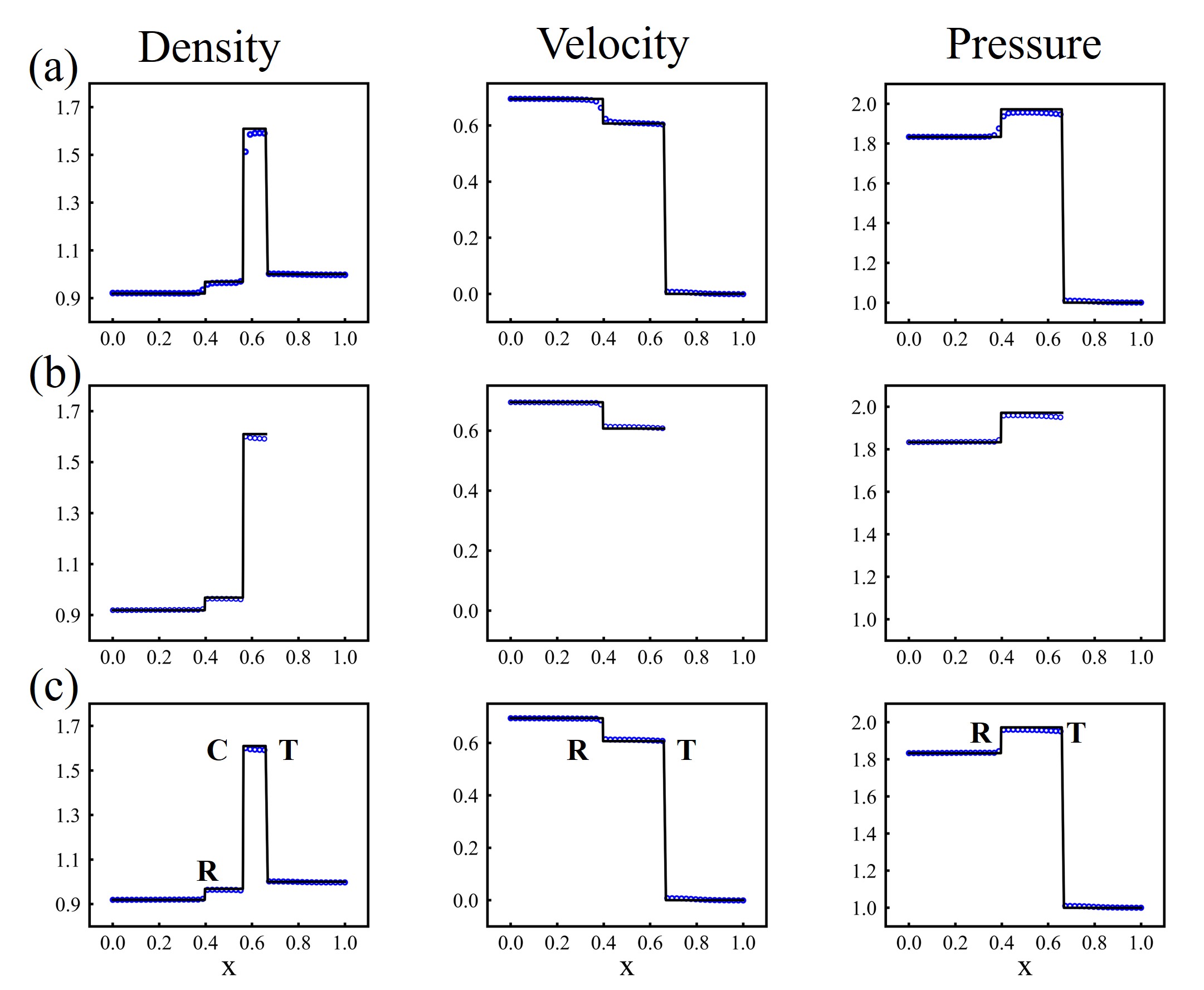}
\caption{The comparison between the neural network predictions (blue circles) and the theoretical solution (black solid line) for density, velocity, and pressure at $t=0.1$ in one-dimensional shock-interface interaction problem. R, C, and T correspond to the reflected shock, contact discontinuity, and transmitted shock, respectively. 
(a) Predictions of Model 1.1 trained by PINNs-WE method across the entire solution domain, which accurately capture the transmitted shock front. (b) Transfer learning is applied in the subdomain post of the transmitted shock based on Model 1.1 with Model 1.2 generating and the reflected shock and contact discontinuity accurately captured. (c) Combine the predictions from Models 1.1 and 1.2 at the transmitted shock to obtain the prediction across the entire solution domain.}
\label{ShockRefraction1D}
\end{figure*}

Other hyper-parameters used in ST-PINNs training processes based on PINNs-WE are displayed in Table \ref{table:Parameter1D}. 
According to Algorithm \ref{algorithm:ST-PINNs}, Model 1.1 is trained using the complete PINNs-WE method, which involves both the RH relation and the global physics conservation constraints, and its predictions are represented by blue circles in FIG. \ref{ShockRefraction1D}(a). 
Model 1.1 can accurately capture the sharp transmitted shock but fails to adequately capture the reflected shock and contact discontinuity, due to the stronger discontinuity intensity of the transmitted shock compared to the reflected shock.

\begin{figure}[!htbp]
\centering
\includegraphics{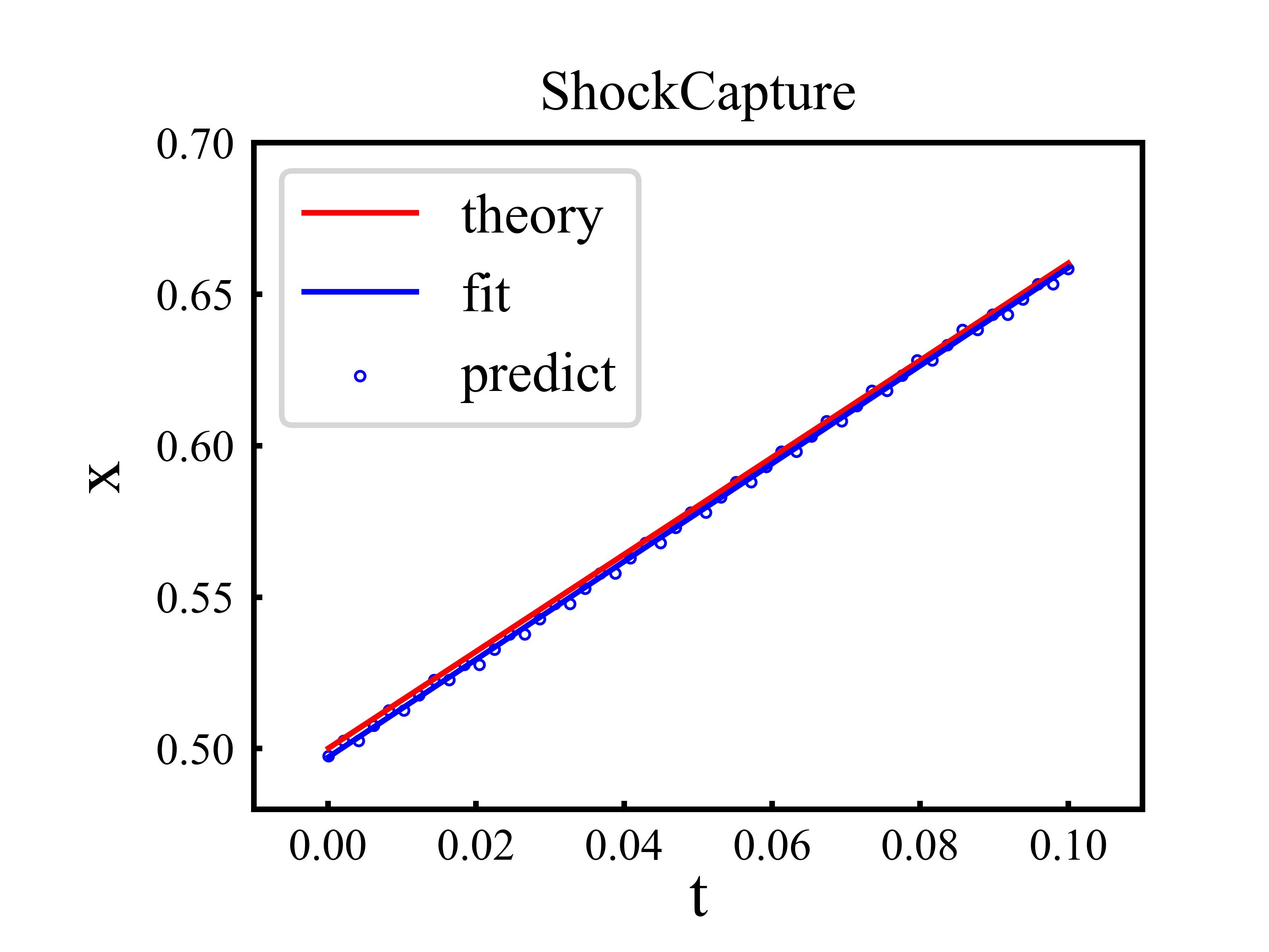}
\caption{The position of the transmitted shock front: 
blue circles is the transmitted shock positions obtained according to Model 1.1, the blue solid line is the fitted positions, and the red solid line represents the theoretical solution.}
\label{fig:Shockcapture}
\end{figure}

After obtaining Model 1.1, the location of the transmitted shock is identified based on the discontinuity strength $I$. 
The time range $t=0.0\sim 0.1$ is then divided into 50 equal intervals and 200 points are uniformly sampled in the spatial range $x=0.0\sim 1.0$ at each time step. 
The intensity of the discontinuity $I$ between adjacent spatial points is calculated based on the pressure difference, and the location with the maximum $I$ is identified as the primary discontinuity. 
The relationship between the discontinuity position $x_{\rm{shock}}$ and time $t$ is then fitted as $x_{\rm{shock}}=1.606\,t+0.5$, as shown in FIG. \ref{fig:Shockcapture}. Furthermore, the average predictions near the transmitted shock front by Model 1.1 are  $\rho=1.590$, $u=0.603$, and $p=1.945$.

The solution domain is then divided into two subdomains from the transmitted shock front. 
The post-shock subdomain contains poorly trained reflected shock front and contact discontinuity. 
Transfer learning in this subdomain based on Model 1.1 results in Model 1.2.
Therein, the boundary conditions are set using the predictions of Model 1.1 in the region near the transmitted shock front.
The predictions of Model 1.2 are represented with blue circles in FIG. \ref{ShockRefraction1D}(b). 

In this step, both the contact discontinuity and the reflected shock are well captured, thus further domain decomposition is unnecessary. 
Finally, the predictions from Models 1.1 and 1.2, corresponding to the pre-shock and post-shock regions respectively are combined at the transmitted shock front, yielding the result for the entire solution domain, as illustrated in FIG. \ref{ShockRefraction1D}(c).  
By comparing FIG. \ref{ShockRefraction1D}(a) and (c), it can be seen that ST-PINNs can effectively improve the predictions by PINNs-WE.

To demonstrate the validity of the ST-PINNs method, a GA-PINNs-based ST-PINNs is applied to solve the same problem as a complementary test. 
The procedures are similar to the PINNs-WE-based ST-PINNs training, thus we omit details here.

The results of PINNs-WE and ST-PINNs based on it, as well as GA-PINNs and its relative ST-PINNs are summarized in Table \ref{table:1D}. 
The results demonstrate that ST-PINNs makes great improvements based on these approaches, and exhibits superior performance in multi-discontinuity problems.
Actually, ST-PINNs provides freedom to choose a model in the first step, readers can use any model to their best knowledge which succeeds in capturing at least one discontinuity.

\begin{table}[!ht]
\centering
\caption{\label{table:1D} $L^2$ relative error of test for one-dimensional shock-interface interaction problem.}
\begin{tabular}{p{1.5cm}p{2.0cm}p{1.5cm}p{1.5cm}p{1.1cm}}
\toprule
Case & Method &  \multicolumn{1}{l}{$L^2 \rho$} & \multicolumn{1}{l}{$L^2 u$} & \multicolumn{1}{l}{$L^2 p$} \\
\hline

\multirow{2}{*}{1.1} 
& PINNs-WE & 1.45\% & 1.11\% & 0.76\% \\
& ST-PINNs & 0.12\% & 0.50\% & 0.21\% \\
\hline

\multirow{2}{*}{1.2} 
& GA-PINNs & 3.38\% & 1.90\% & 0.98\% \\
& ST-PINNs & 0.46\% & 0.82\% & 0.41\% \\
\hline
\end{tabular}
\end{table}

\subsection{Quasi one-dimensional planar shock-interface interaction}\label{section4_2}

\begin{figure}[H]
\centering
\includegraphics{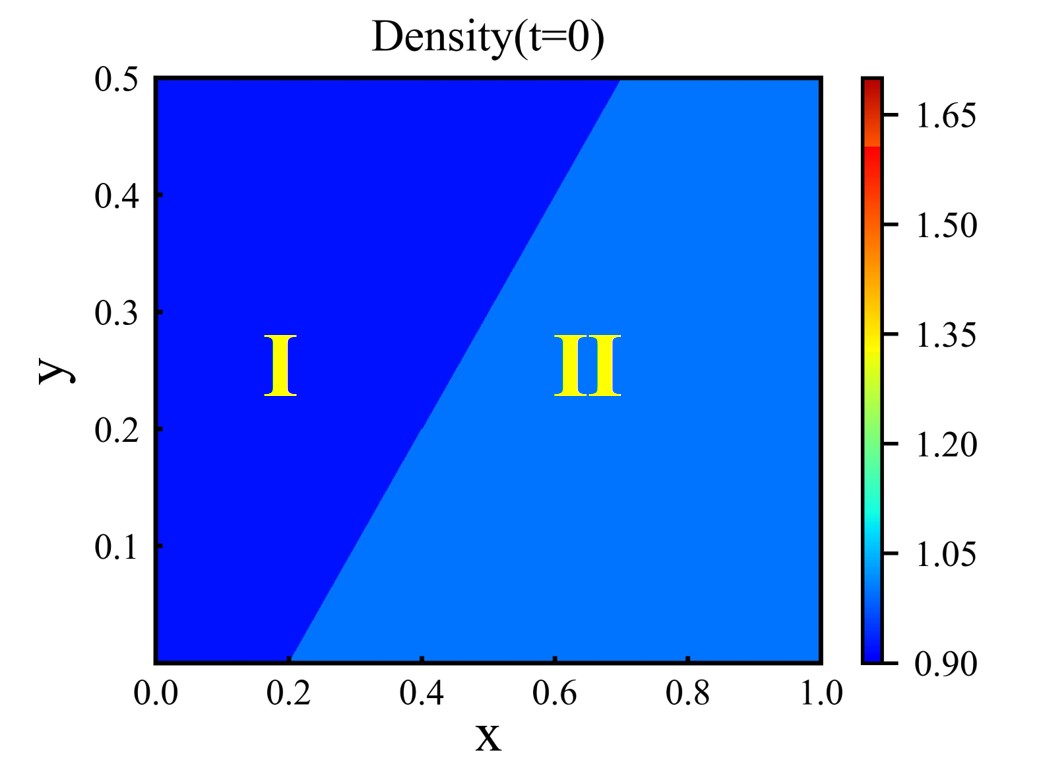}
\caption{Initial conditions of the mass density $\rho$ for the quasi one-dimensional planar shock-interface interaction problem. $\rm{Region \ \uppercase\expandafter{\romannumeral 1}}$ is the light medium compressed by the incident shock, and $\rm{region \ \uppercase\expandafter{\romannumeral 2}}$ is the heavy medium.}
\label{fig:ShcokRefractionQ1D_density}
\end{figure}

We apply ST-PINNs to a quasi one-dimensional problem, which is a two-dimensional extension of the one-dimensional shock-interface interaction problem. The entire solution domain includes a transmitted shock, a contact discontinuity, and a reflected shock.
The solution domain is set to $x\times y\times t=[0.0,1.0]\times [0.0,0.5]\times[0.0,0.1]$, with the initial conditions given by Eq. \ref{eq:ShockRefractionQ1D}.  

\begin{equation}
    \rho,u,v,p=\begin{cases}
    (0.919,\,\,0.491,\,\,-0.491,\,\,1.833),  & \rm{ if \ region \ \uppercase\expandafter{\romannumeral1}}; \\
    (1.0,\,\,0.0,\,\,0.0,\,\,1.0),      & \rm{ if \ region \ \uppercase\expandafter{\romannumeral2}}. \\
    \end{cases}
    \label{eq:ShockRefractionQ1D}
\end{equation}
where \rm{region \uppercase\expandafter{\romannumeral1}} is $x-0.2<y$, \rm{region \uppercase\expandafter{\romannumeral2}} is $x-0.2>y$, as shown in FIG. \ref{fig:ShcokRefractionQ1D_density}.

\begin{table}
\centering
\caption{\label{table:ParameterQ1D} Hyper-parameters of ST-PINNs based on generalized GA-PINNs method. Stage 1 refers to the epochs from 1 to 30000, Stage 2 refers to the epochs from 30001 to 50000, Stage 3 refers to the epochs from 50001 to 60000.}
\begin{tabular}
{p{4.0cm}p{2.5cm}p{1.5cm}p{0.8cm}}
    \toprule  
     Parameters & Model 2.1 & Model 2.2 \\
    \hline
     Number of layers  & 10 & 10 \\
     Neurons & 100 & 100     \\
     Residual Points & 48000 & 40000 \\
     Initial points & 5000 & 5000 \\
     Boundary Points &NAN & 2000\\
     Epoch & 60000 & 60000 \\
     Learning rate(Stage 1) & 1e-3 & 1e-5\\
     Learning rate(Stage 2) & 1e-5 & 1e-6 \\
     Learning rate(Stage 3) & 1e-6 & 1e-7\\
     $\omega_{\rm{F}}$ & 1 & 1  \\
     $\omega_{\rm{IC}}$ & 10 & 10\\
     $\omega_{\rm{BC}}$ & NAN & 10\\
    \hline
\end{tabular}
\end{table}

In this case, a better prediction comes from ST-PINNs training based on two-dimensional generalized GA-PINNs, which is discussed in detail in this part.
Meanwhile, results from ST-PINNs based on a simplified PINNs-WE like method are shown as a complementary test.

\begin{figure*}[htbp]
\centering
\includegraphics[scale=1.0]{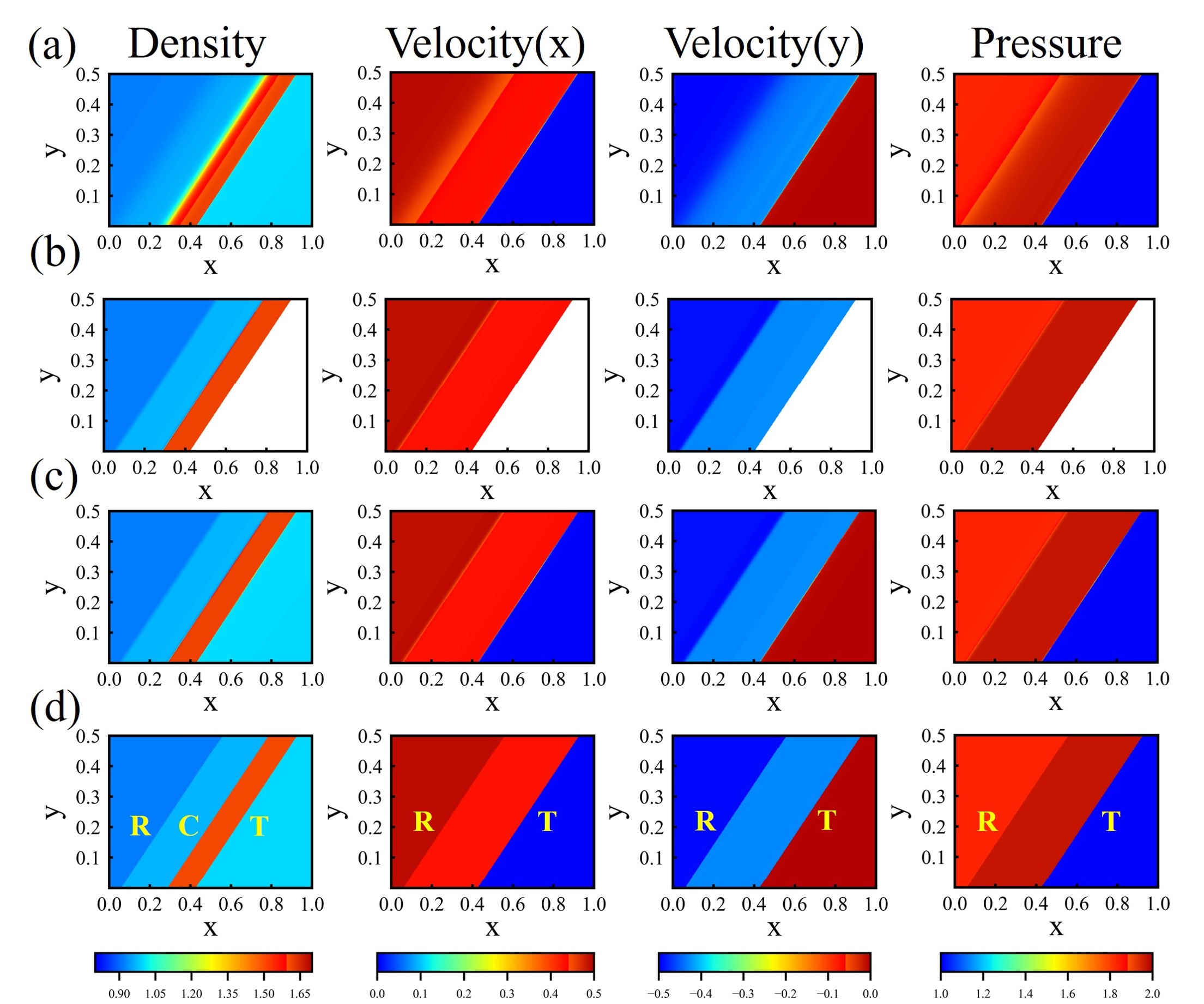}
\caption{Results of the quasi-one-dimensional planar shock-interface interaction problem at $t=0.1$ solved by ST-PINNs based on GA-PINNs. R, C, and T correspond to the reflected shock, contact discontinuity, and transmitted shock, respectively.   
(a) Results solved by the GA-PINNs method, generating Model 2.1, with the transmitted shock well-captured. 
(b) 
Transfer learning based on Model 2.1 is applied to the subdomain post the transmitted shock, generating Neural Network Model 2.2, with the reflected shock and contact discontinuity well captured. 
(c) Predictions at both sides of the transmitted shock front provided by Models 2.1 and 2.2 respectively are combined to obtain the results across the entire solution domain. 
(d) Theoretical solutions.}
\label{fig:ShcokRefractionQ1D}
\end{figure*}

The hyper-parameters of the GA-PINNs based ST-PINNs model are shown in Table \ref{table:ParameterQ1D}.
First, a two-dimensional generalized GA-PINNs method is used to train Model 2.1,  with a local weight 
\begin{equation}
\lambda =\frac{1}{\alpha_{\rho}|\nabla\rho|^{\beta_{\rho}}+\alpha_{u}|\nabla \vec{u}|^{\beta_{u}}+\alpha_{p}|\nabla p|^{\beta_{p}}+1  } . \label{eq:lambda_shock2DR}
\end{equation}
where $\alpha$'s and $\beta$'s are set as $\alpha_{\rho}=\alpha_{u}=\alpha_{p}=\alpha=0.4$, $\beta_{\rho}=\beta_{u}=\beta_{p}=\beta=0.3$.
The predictions of Model 2.1 is shown in FIG. \ref{fig:ShcokRefractionQ1D}(a). 
During this process, the transmitted shock is well-captured but other discontinuities are poorly-captured. 

To capture the position of the transmitted shock, we can use the intrinsic physical characteristic that transmitted shock is planar. Thus, the position of the transmitted shock front can be determined via two points at a specific time. 
The time range $t=0.0\sim 0.1$ is then divided into 100 equal intervals and at each time step, 200 points are uniformly sampled along the $x$-direction at $y=0.0$ and $y=0.5$.
The discontinuity intensity $I$ between adjacent spatial grid points at each time step is calculated based on the pressure difference, and the location with the maximum $I$ is identified as the primary discontinuity at $y=0.0$ and $y=0.5$. 
The position of the shock front is fitted as $x_1 = 2.228\, t + 0.20$ for $y = 0.0$ and $x_{2} = 2.232\, t + 0.70$ for $y = 0.5$, enabling the determination of the transmitted shock position at each time step.
The predictions in the region near the transmitted shock front are $\rho=1.618$, $u=0.429$, $v=-0.426$, and $p=1.968$.

Subsequently, the entire solution domain is divided into two subdomains from the fitted transmitted shock front. 
The post-shock subdomain contains the poorly-captured reflected shock and the contact discontinuity. 
Model 2.2 is obtained by transfer learning for this subdomain based on Model 2.1,
with boundary conditions taken from the predictions of Model 2.1 in the region near the transmitted shock front. 
The predictions of Model 2.2 are shown in FIG. \ref{fig:ShcokRefractionQ1D}(b). This result shows that the reflected shock and the contact discontinuity are well trained.

Finally, the predictions from Models 2.1 and 2.2, corresponding to the pre-shock and post-shock regions, are combined at the transmitted shock front to obtain the final result for the entire solution domain, as shown in FIG. \ref{fig:ShcokRefractionQ1D}(c).  The results from ST-PINNs are consistent with the theoretical solution shown in FIG. \ref{fig:ShcokRefractionQ1D}(d).

In this complementary test, a simplified PINNs-WE like method is applied as a base model, with a modified version of Eq. (\ref{eq:lambda}) as 
\begin{equation}
 \lambda =\frac{1}{k(\sqrt{(\partial_x u)^2+(\partial_y v)^2}+1}, \label{eq:simplified_lambda_PINNs-WE}
\end{equation}
and $k=0.1$. 
Since this simplified model with only the modified local weight Eq. (\ref{eq:simplified_lambda_PINNs-WE}) already yields sufficiently accurate primary discontinuity in the first training step, we then neglect the RH relations and global conservation constraints in this case to streamline the training process.
The results still indicate that the ST-PINNs method offers improvements over the baseline model.

The prediction errors of different models relative to the theoretical solution are evaluated using $x \times y=400 \times 200$ sample points at the end ($t = 0.1$). 
The errors of the GA-PINNs model, which is actually the result shown in FIG.\ref{fig:ShcokRefractionQ1D}(a), are $4.62\%$ for density, $4.03\%$ for velocity in the $x$-direction, $4.05\%$ for velocity in the $y$-direction, and $2.12\%$ for pressure. 
After training with ST-PINNs, these errors are reduced to $1.80\%$, $3.49\%$, $3.50\%$, and $1.85\%$, demonstrating an improvement in prediction accuracy. 
The results of ST-PINNs based on GA-PINNs and simplified PINNs-WE like methods are summarized in Table \ref{table:Q1D}. 
The results demonstrate that ST-PINNs exhibits superior performance in multi-discontinuity problems.

\begin{table}[!ht]
\centering
\caption{\label{table:Q1D} $L^2$ relative error of test for quasi-one-dimensional planar shock-interface interaction problem.}
\begin{tabular}{p{1.0cm}p{2.3cm}p{1.3cm}p{1.3cm}p{1.3cm}p{1.1cm}}
\toprule
Case & Method &  \multicolumn{1}{l}{$L^2 \rho$} & \multicolumn{1}{l}{$L^2 u$} & \multicolumn{1}{l}{$L^2 v$} &\multicolumn{1}{l}{$L^2 p$} \\
\hline

\multirow{2}{*}{2.1} 
&PINNs-WE like &$5.01\%$ & $4.78\%$&$4.80\%$ &$2.41\%$\\
&ST-PINNs & $2.24\%$ &$4.32\%$ &$4.30\%$ &$2.19\%$\\
\hline

\multirow{2}{*}{2.2} 
& GA-PINNs & 4.62\% & 4.03\% & 4.05\% & 2.12\%\\ 
& ST-PINNS & 1.80\% & 3.49\% & 3.50\%  & 1.85\%\\
\hline
\end{tabular}
\end{table}

\subsection{Two-dimensional unsteady planar shock refraction}\label{section4_3}

\begin{figure}[!htbp]
\centering
\includegraphics{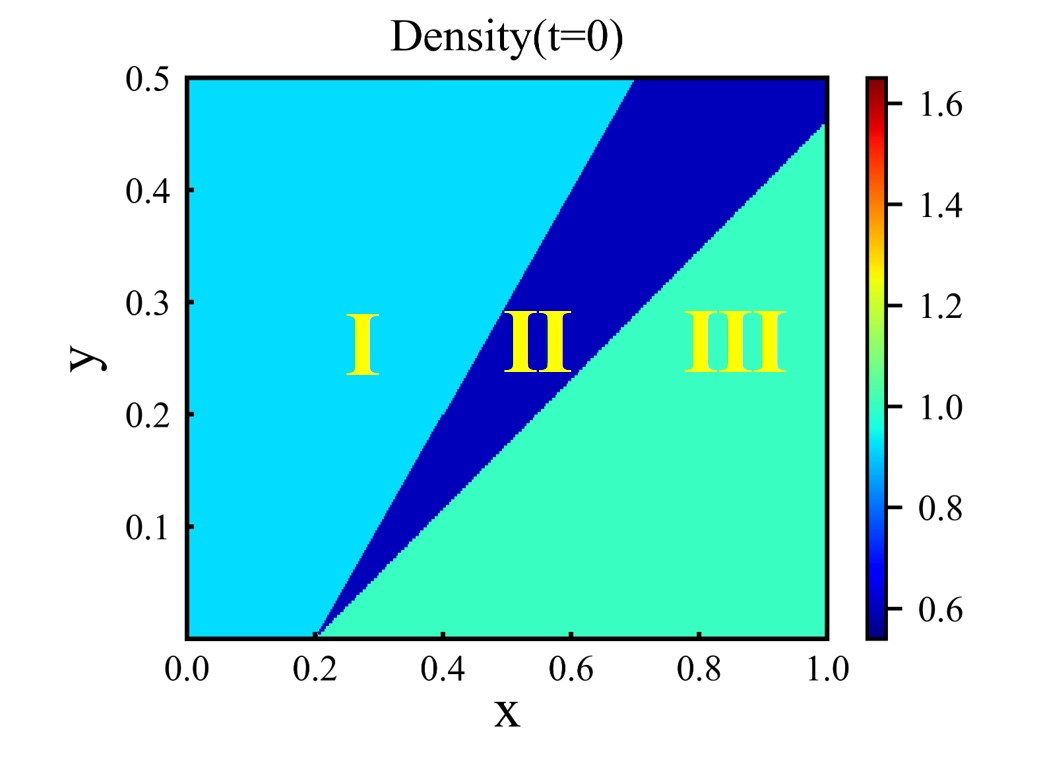}
\caption{Initial conditions of the mass density $\rho$ for the two-dimensional planar shock refraction problem. $\rm{Region 
\ \uppercase\expandafter{\romannumeral1}}$ is the compressed light medium downstream of the incident shock, $\rm{region \ \uppercase\expandafter{\romannumeral2}}$ is the uncompressed light medium upstream of the incident shock, and $\rm{region \ \uppercase\expandafter{\romannumeral3}}$ is the heavy medium upstream of the shock.}
\label{fig:ShcokRefraction2D_density}
\end{figure}

\begin{figure*}[htbp]
\centering
\includegraphics[scale=1.0]{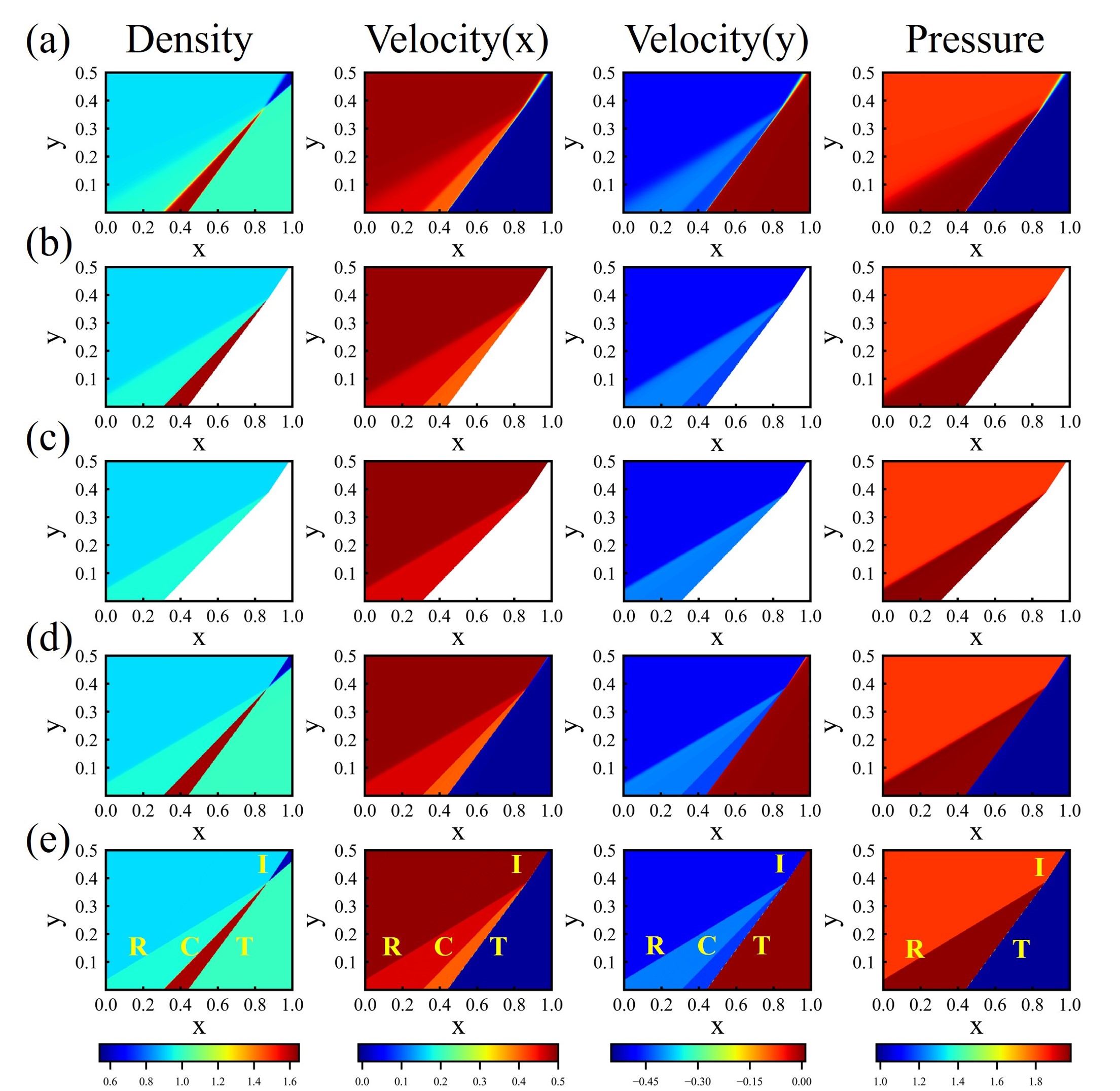}
\caption{Step-by-step result plots by ST-PINNs for two-dimensional planar shock refraction problem. Four columns show the solutions for density, $x$-direction velocity, $y$-direction velocity, and pressure at time $t=0.1$. I, R, C, and T correspond to the incident shock, reflected shock, contact discontinuity, and transmitted shock, respectively.  
(a) The training results by GA-PINNs method with the local weight Eq. (\ref{eq:lambda_shock2DR}), with the transmitted shock well-trained and Model 3.1 generated. (b) Results of transfer learning based on Model 3.1 applied in the subdomain post the transmitted shock and incident shock, with incident shock, trained transmitted shock and their post physics quantities as boundary conditions. Model 3.2 is generated and the contact discontinuity is well trained. (c) Results of transfer learning based on Model 3.2 applied in the subdomain post the contact discontinuity and the incident shock. 
Model 3.3 is generated and the reflected shock is well-trained. (d) Combine predictions of Models 3.1, 3.2, and 3.3 in appropriate subdomains to obtain the results across the entire solution domain.
(e) Numerical solution by JAX-FLUIDS.}
\label{fig:ShockRefraction2D}
\end{figure*}

In this part, we introduce the application of ST-PINNs to solve the two-dimensional unsteady  planar shock refraction problem. 
When the incident shock propagates from a light to a heavy medium across an inclined interface, it generates a transmitted shock, a reflected shock, and an interface deflection at a certain angle.  
In this case, The solution domain is set to $x \times y \times t=[0.0,\,1.0] \times [0.0,\,0.5] \times [0.0,\, 0.1]$, with the initial conditions given by Eq. (\ref{eq:shockrefraction2d}).

\begin{equation}
    \rho,u,v,p=\begin{cases}
    (0.919,\,\,0.491,\,\,-0.491,\,\,1.833),  & \mathrm{ if \ region \ \uppercase\expandafter{\romannumeral1}}; \\
    (0.693,\,\,0.0,\,\,0.0,\,\,1.0),      & \mathrm{ if \ region \ \uppercase\expandafter{\romannumeral2}}; \\
    (1.0,\,\,0.0,\,\,0.0,\,\,1.0),      & \mathrm{ if \ region \ \uppercase\expandafter{\romannumeral3}}. \\
    \end{cases}
    \label{eq:shockrefraction2d}
\end{equation}
where $\rm{region 
\ \uppercase\expandafter{\romannumeral1}}$ is $x-0.2<y$, $\rm{region \ \uppercase\expandafter{\romannumeral2}}$ is $0.577\,x-0.115<y\le x-0.2$, $\rm{region \ \uppercase\expandafter{\romannumeral3}}$ is $y < 0.577\,x - 0.115$, as shown in FIG. \ref{fig:ShcokRefraction2D_density}.

\begin{table}
\centering
\caption{\label{table:Parameter2D} Hyper-parameters of ST-PINNs based on generalized GA-PINNs method. Stage 1 refers to the epochs from 1 to 30000, Stage 2 refers to the epochs from 30001 to 50000, Stage 3 refers to the epochs from 50001 to 60000.}
\begin{tabular}
{p{3.2cm}p{1.8cm}p{1.8cm}p{1.5cm}p{0.4cm}}
    \toprule 
    
     Parameters & Model 3.1 & Model 3.2 & Model 3.3\\
    
    \hline
     Number of layers  & 10 & 10  & 10\\
     Neurons & 100 & 100     & 100\\
     Residual Points & 48000 & 44000  & 44000\\
     Initial points & 5000 & 5000 & 5000\\
     Boundary Points &NAN & 4000 & 4000\\
     Epoch & 60000 & 60000 & 60000\\
     Learning rate(Stage 1) & 1e-3 & 1e-5 & 1e-5\\
     Learning rate(Stage 2) & 1e-5 & 1e-6 & 1e-6\\
     Learning rate(Stage 3) & 1e-6 & 1e-7 & 1e-7\\
     $\omega_{\rm{F}}$ & 1 & 1 & 1 \\
     $\omega_{\rm{IC}}$ & 10 & 10 & 10\\
     $\omega_{\rm{BC}}$ & NAN & 10 & 10\\
    
    \hline
\end{tabular}
\end{table}

As in Section \ref{section4_2}, we have a better prediction from ST-PINNs training based on two-dimensional generalized GA-PINNs.
Meanwhile, results from ST-PINNs based on a simplified PINNs-WE like method is shown as a complementary test.

The hyper-parameters for ST-PINNs based on GA-PINNs are shown in Table \ref{table:Parameter2D}. First, the two-dimensional generalized GA-PINNs method is used to solve the problem throughout solution domain, with a local weight as shown in Eq. \ref{eq:lambda_shock2DR} and $\alpha_{\rho}=\alpha_{u}=\alpha_{p}=\alpha=0.2$, $\beta_{\rho}=\beta_{u}=\beta_{p}=\beta=0.5$.

Model 3.1 is then generated, with its predictions shown in FIG. \ref{fig:ShockRefraction2D}(a).
During the training process, the transmitted shock is captured well, while the reflected shock, contact discontinuity, and incident shock are poorly-captured.

For the two-dimensional unsteady planar shock refraction problem, the prior information, in particular the intersection point between the incident shock and the inclined interface at any given time, is used to reduce the training difficulty and enhance the accuracy of the solution. 
Since the transmitted shock is planar, we can determine the position of the transmitted shock front using the intersection point and another point on the transmitted shock front. 
To capture the point on the transmitted shock front, the time range $t=0.0\sim 0.1$ is divided into 100 equal intervals, and 200 points are uniformly sampled along the $x$-direction at $y=0.0$ at each time step.
Similarly as in previous cases, the intensity of the discontinuity $I$ between adjacent spatial grid points at each time step is calculated based on the pressure difference, and the location with the maximum $I$ is identified as the primary discontinuity at $y=0$. 
The position of the transmitted shock front is then fitted to $x_{\rm{shock}}=2.370\,t+0.2$ for $y = 0.0$.
The average predictions in the region near the transmitted shock front are $\rho_{\rm{NN}}=1.617$, $u_{\rm{NN}}=0.401$, $v_{\rm{NN}}=-0.440$, and $p_{\rm{NN}}=1.962$.

Subsequently, the entire solution domain is divided into two subdomains according to the transmitted shock front and the incident shock front. 
The post-shock subdomain contains the poorly trained reflected shock, contact discontinuity, and incident shock front. 
{\color{black} Note that the position of the incident shock front is a known prior information which can be directly used in the following training.}
Transfer learning is applied to this subdomain based on Model 3.1, with boundary conditions taken from the predictions of Model 3.1 in the region near the transmitted shock front and the known value in the region near incident shock front.

After transfer learning, Model 3.2 is obtained, and its predictions are shown in FIG. \ref{fig:ShockRefraction2D}(b). In this step, the contact discontinuity and the incident shock front are well trained, but the reflected shock remains poorly trained. 
Therefore, further domain separation is performed at the contact discontinuity interface. 
The position of the contact discontinuity is determined using a method similar to that for the transmitted shock front. 
The position of the contact discontinuity interface is fitted as $x_{\rm{contact}}=1.05\,t+0.2$ for $y = 0.0$. 
The average predictions in the region near the contact discontinuity interface are $\rho_{\rm{NN}}=0.960,u_{\rm{NN}}=0.457,v_{\rm{NN}}=-0.409$ and $p_{\rm{NN}}=1.968$.
Transfer learning is again applied to generate Model 3.3 based on Model 3.2.
The results are shown in FIG. \ref{fig:ShockRefraction2D}(c), where the reflected shock is well-trained.
Finally, the predictions from Model 3.1, 3.2, and 3.3 are combined across their respective subdomains to reconstruct the predictions of the entire solution domain, as shown in Fig. \ref{fig:ShockRefraction2D}(d). In this step, the fronts of the transmitted and reflected shocks, and the contact discontinuity are all accurately captured. 
The ST-PINNs results, which effectively capture sharp discontinuities, are perfectly consistent with the numerical solutions from JAX-FLUIDS \cite{JAX-Fluids}, as shown in FIG. \ref{fig:ShockRefraction2D}(e). 
These conventional numerical solutions are obtained using the finite volume method, with detailed settings for JAX-FLUIDS are provided in Appendix. \ref{Appendix2}.

\begin{table}[!ht]
\centering
\caption{\label{table:2D} $L^2$ relative error of test for two-dimensional planar shock refraction problem.}
\begin{tabular}{p{1.0cm}p{2.3cm}p{1.3cm}p{1.3cm}p{1.3cm}p{1.1cm}}
\toprule
Case & Method &  \multicolumn{1}{l}{$L^2 \rho$} & \multicolumn{1}{l}{$L^2 u$} & \multicolumn{1}{l}{$L^2 v$} &\multicolumn{1}{l}{$L^2 p$} \\
\hline

\multirow{2}{*}{3.1} 
&PINNs-WE like &$3.80\%$ & $6.55\%$&$7.22\%$ &$3.46\%$\\
&ST-PINNs& $1.48\%$ &$1.32\%$ &$1.82\%$ &$0.83\%$\\
\hline

\multirow{2}{*}{3.2} 
& GA-PINNs & 3.55\% & 6.03\% & 6.66\% & 3.28\%\\
& ST-PINNS & 1.20\% & 1.33\% & 1.84\% & 0.77\%\\
\hline
\end{tabular}
\end{table}

The prediction errors of different models relative to the numerical solutions provided by JAX-FLUIDS are evaluated using $x \times y=400 \times 200$ sample points at the ending time ($t = 0.1$). 
Similar to the analysis in Section \ref{section4_2}, the results of ST-PINNs based on GA-PINNs and simplified PINNs-WE-like methods are summarized in Table \ref{table:2D}.
As shown in Table \ref{table:2D}, once the primary discontinuity is accurately captured in the initial training step, ST-PINNs consistently improve upon their corresponding baseline models. This demonstrates the superior performance of ST-PINNs in handling problems involving multiple discontinuities.

\section{DISCUSSION AND CONCLUSION}\label{section5}

This paper proposes the ST-PINNs method for solving hydrodynamic problems with multiple discontinuities.
The effectiveness of the ST-PINNs is validated through its application to the one-dimensional shock-interface interaction problem, the quasi-one-dimensional planar shock-interface interaction problem, and the two-dimensional planar shock refraction problem. 
ST-PINNs, which solves each discontinuity sequentially and applying transfer learning during training, can successfully capture the discontinuities and predict the correct physical quantities in these problems.
This method significantly improves accuracy over the standard PINNs and its other variations.
In the cases studied in this paper, once the base model in the first training step captures at least one discontinuity with a certain precision, ST-PINNs then can always improve the predictive performance.
To the best of our knowledge, this is the first time that a variant of PINNs has been used to solve two-dimensional unsteady planar shock refraction problems, providing a valuable reference for extending PINNs-based approaches to more complex shock-interface interaction problems.
Furthermore, ST-PINNs is not only effective for hydrodynamic problems but can also be extended to address problems involving multiple discontinuities in other research fields.

Although ST-PINNs features a more complex algorithmic workflow, it significantly improves training performance in problems with multiple discontinuities. 
We expect ST-PINNs to be applied to more complex shock-interface interaction problems and flow field reconstruction. In flow field reconstruction, additional physical information, such as density and velocity gradients, along with sparse data, can be used to improve prediction accuracy. We believe the inclusion of sparse data would further enhance the predictive performance of ST-PINNs.

To simulate infinitely long interfaces, no boundary conditions have been applied in the numerical experiments presented in this paper. In the future, we expect the application of ST-PINNs to problems with actual boundary conditions, such as no-slip wall boundaries. Currently, we have considered only one-dimensional and two-dimensional problems. In this work, we have fully recognized the challenges posed by gradient pathologies on PINNs, and we aim to refine the algorithms to better address these issues.
We hope to further apply this method to problems with boundary conditions and expand ST-PINNs to three-dimensional scenarios.

\begin{acknowledgments}
This work was supported by the National Natural Science Foundation of China (Grant No. U2330401(NSAF)).
\end{acknowledgments}

\appendix

\section{Settings of JAX-FLUIDS in Two-Dimensional Unsteady Planar Shock Refraction Problem}\label{Appendix2}

This section details the specific setup of JAX-FLUIDS for solving the two-dimensional unsteady planar shock refraction problem.
Since our goal is to simulate the interaction of a shock with an infinitely long inclined interface, while traditional numerical simulations require boundary conditions, we expand the range of the solution domain to avoid the influence of boundary conditions on the simulation results. 
The solution domain is set to $x \times y \times t=[-0.7,1.3] \times [-0.386,1.0] \times[0.0,0.2]$, with a grid resolution of $x \times y=3000 \times 3000$, equivalent to the two-dimensional unsteady planar shock refraction problem described in Section. \ref{section4_3}.
The boundary conditions are set to ``ZEROGRADIENT'' with a CFL number of 0.5. The solver is HLLC, the spatial discretization scheme is WENO-5, and the time integration scheme is RK3.

\section{Test of ST-PINNs}

To demonstrate the effectiveness and stability of ST-PINNs, this paper adopts varying parameters in the first training step for numerical examples and subsequently employs ST-PINNs to enhance training precision. 
The test results of Section \ref{section4_1} are shown in Table. \ref{test4.1}. 
The test results of Section \ref{section4_2} and Section \ref{section4_3} are shown in Table. \ref{test4.24.3}.

\begin{table}[!ht]
\centering
\caption{\label{test4.1} $L^2$ relative error of test for Section \ref{section4_1}. 
PINNs-WE, GA-PINNs, and their corresponding ST-PINNs are denoted by M1, M2, and M3 respectively.}
\begin{tabular}{p{1.0cm}p{1.3cm}p{0.8cm}p{0.8cm}p{1.5cm}p{1.5cm}p{1.1cm}}
\toprule
Test & Method & \multicolumn{2}{c}{k} &\multicolumn{1}{l}{$L^2 \rho$} & \multicolumn{1}{l}{$L^2 u$} & \multicolumn{1}{l}{$L^2 p$} \\
\hline

\multirow{2}{*}{1.1} 
& M1 & \multicolumn{2}{c}{0.6} & 1.45\% & 1.11\% & 0.76\% \\
& M3 & \multicolumn{2}{c}{0.6} & 0.12\% & 0.50\% & 0.21\% \\
\hline

\multirow{2}{*}{1.2} 
& M1 & \multicolumn{2}{c}{0.1} & 2.38\% & 2.56\% & 1.75\% \\
& M3 & \multicolumn{2}{c}{0.1} & 0.64\% & 1.21\% & 0.72\% \\
\hline

\multirow{2}{*}{1.3} 
& M1 & \multicolumn{2}{c}{0.8} & 3.92\% & 2.23\% & 1.34\% \\
& M3 & \multicolumn{2}{c}{0.8} & 0.62\% & 1.17\% & 0.66\% \\

\hline
\hline
 &  & $\alpha$ & $\beta$ &
 &  &  \\
\hline

\multirow{2}{*}{1.4} 
& M2 & 0.2& 0.1& 3.64\% & 1.91\% & 0.98\% \\
& M3 & 0.2& 0.1& 0.49\% & 1.01\% & 0.51\% \\
\hline

\multirow{2}{*}{1.5} 
& M2 &0.2 & 0.5& 3.37\% & 2.08\% & 1.08\% \\
& M3 & 0.2& 0.5& 0.47\% & 0.82\% & 0.41\% \\
\hline
\multirow{2}{*}{1.6}
& M2 & 0.4& 0.3& 3.38\% & 1.90\% & 0.98\% \\
& M3 & 0.4& 0.3& 0.46\% & 0.82\% & 0.41\% \\
\hline
\end{tabular}
\end{table}

\begin{table}[!ht]
\centering
\caption{\label{test4.24.3} $L^2$ relative error of test for Section. \ref{section4_2} and Section. \ref{section4_3}. GA-PINNs and ST-PINNs are denoted by M2, and M3 respectively.}
\begin{tabular}{p{0.8cm}p{1.3cm}p{0.6cm}p{0.6cm}p{1.1cm}p{1.1cm}p{1.1cm}p{1.1cm}}
\toprule
Test & Method & $\alpha$& $\beta$ &\multicolumn{1}{l}{$L^2 \rho$} & \multicolumn{1}{l}{$L^2 u$} & \multicolumn{1}{l}{$L^2 v$} & \multicolumn{1}{l}{$L^2 p$} \\
\hline

\multirow{2}{*}{2.1} 
& M2 &0.4& 0.3& 4.62\% & 4.03\% & 4.05\% & 2.12\% \\
& M3 &0.4& 0.3 & 1.80\% & 3.49\% & 3.50\% & 1.85\%\\
\hline

\multirow{2}{*}{2.2} 
& M2 &0.2& 0.5 & 4.80\% & 4.41\% & 4.42\% & 2.36\%\\
& M3 &0.2& 0.5 & 2.04\% & 3.92\% & 3.93\% & 2.10\% \\
\hline

\multirow{2}{*}{2.3} 
& M2 &0.2& 0.1 & 4.99\% & 4.22\% & 4.23\% & 2.16\%\\ 
& M3 &0.2& 0.1 & 1.65\% & 3.15\% & 3.15\%  & 1.59\%\\

\hline

\multirow{2}{*}{3.1} 
& M2 &0.2& 0.5 & 3.55\% & 6.03\% & 6.66\% & 3.28\%\\
& M3 &0.2& 0.5 & 1.20\% & 1.33\% & 1.84\% & 0.77\%\\
\hline

\multirow{2}{*}{3.2} 
& M2 &0.4& 0.3 & 3.85\% & 6.50\% & 7.14\% & 3.60\%\\
& M3 &0.4& 0.3& 1.43\% & 1.94\% & 2.38\% & 1.12\%\\
\hline
\multirow{2}{*}{3.3}
& M2 &0.2& 0.1& 3.67\% & 5.95\% & 6.64\% & 3.24\%\\
& M3 &0.2& 0.1 & 1.65\% & 2.24\% & 2.75\% & 1.37\%\\
\hline
\end{tabular}
\end{table}

\section*{References}
\bibliography{ref.bib}

\end{document}